\documentstyle[psfig,12pt]{article}
\newcommand{\Yes}{\rule{1.0in}{0.02in}}
\newcommand{\Yesm}{\hspace{-2em}\rule{1.3in}{0.02in}}
\newcounter{labelflag} 

\newcommand{\Label}[1]{
                       \ifnum\thelabelflag=1
                          \ifmmode
                             \makebox[0in][l]{\qquad\fbox{\rm#1}}
                          \else
                             \marginpar{\vspace{0.7\baselineskip}
                                        \hspace{-1.1\textwidth}
                                        \fbox{\rm#1}}
                          \fi
                       \fi
                       \label{#1}
                      }
\newcommand{\be}{\begin{equation}}
\newcommand{\ee}{\end{equation}}

\begin{document}       
                      

\begin{center}
{\Large\bf Data Sonification and Sound Visualization}
\end{center}
   
\noindent
Hans G.\ Kaper \\
\hspace*{2em}
\textit{
Mathematics and Computer Science Division,
Argonne National Laboratory
} \\
Sever Tipei \\
\hspace*{2em}
\textit{
School of Music,
University of Illinois
} \\
Elizabeth Wiebel \\
\hspace*{2em}
\textit{
Mathematics and Computer Science Division,
Argonne National Laboratory
}

\medskip

\begin{abstract}
This article describes a collaborative project
between researchers in the Mathematics and Computer
Science Division at Argonne National Laboratory
and the Computer Music Project of the University
of Illinois at Urbana-Champaign.
The project focuses on the use of sound for the
exploration and analysis of complex data sets
in scientific computing.
The article addresses digital sound synthesis
in the context of DIASS (Digital Instrument
for Additive Sound Synthesis) and sound
visualization in a virtual-reality environment
by means of M4CAVE.
It describes the procedures and preliminary results
of some experiments in scientific sonification
and sound visualization.
\end{abstract}


\medskip

\noindent
While most computational scientists routinely use
visual imaging techniques to explore and analyze
large data sets, they tend to be much less familiar
with the use of sound.
Yet, sound signals carry significant amounts of
information and can be used advantageously to
increase the bandwidth of the human/computer
interface.
The project described in this article focuses on
scientific sonification---the faithful rendering
of scientific data in sounds---and the visualization
of sounds in a virtual-reality environment.
The project, which grew out of an effort to apply
the latest supercomputing technology to
the process of music composition (see Box~1),
is a joint collaboration between
Argonne National Laboratory (ANL, Mathematics and
Computer Science Division) and the University of
Illinois at Urbana-Champaign (UIUC, Computer Music
Project).

Digital sound synthesis is addressed in Section~1;
the discussion centers on DIASS
(Digital Instrument for Additive Sound Synthesis).
Section~2 describes some experiments in
scientific sonification.
Sound visualization in a virtual-reality (VR)
environment is discussed in Section~3;
here, the main tool is M4CAVE, a program to
visualize sounds from a score file.
Section~4 contains some general observations
about the project.

\section{Digital Sound Synthesis}
Digital sound synthesis is a way to generate
a stream of numbers representing the sampled
values of an audio waveform.
To realize the sounds, one sends these samples
through a digital-to-analog converter (DAC),
which converts the numbers to a continuously
varying voltage that can be amplified and sent to
a loudspeaker.

One way of viewing the digital sound-synthesis process
is to imagine a computer program that calculates
the sample values according to a mathematical formula
and sends those samples, one after the other, to the DAC.
All the calculations are carried out by a program,
which can be changed in arbitrary ways by the user.
From this point of view, digital synthesis is the same
as software synthesis.
Software synthesis contrasts with hardware synthesis,
where the calculations are carried out in special
circuitry.
Hardware synthesis has the advantage of high-speed
operation but lacks the flexibility of software
synthesis.
Software synthesis is the technique of choice
if one wishes to develop an instrument for
data sonification.

With software synthesis, one can indeed realize any
imaginable sound---provided one has the time
to wait for the results.
With a sampling rate of 44,100 samples per second
the time available per sample is only 20 microseconds,
too short for real-time synthesis of reasonably complex sounds.
For this reason, most of today's synthesis programs
generate a sound file, which is then played through a DAC.
But data sonification in real time may become feasible
on tomorrow's high-performance computing architectures.
Our research effort focuses on the development of
a flexible and powerful digital instrument for
scientific sonification and on finding optimal ways
to convey information through the medium of sound.

\subsection{DIASS -- A Digital Instrument}

\begin{figure}[htbp]
\hspace*{-0.0in}  
\centering{\psfig{file=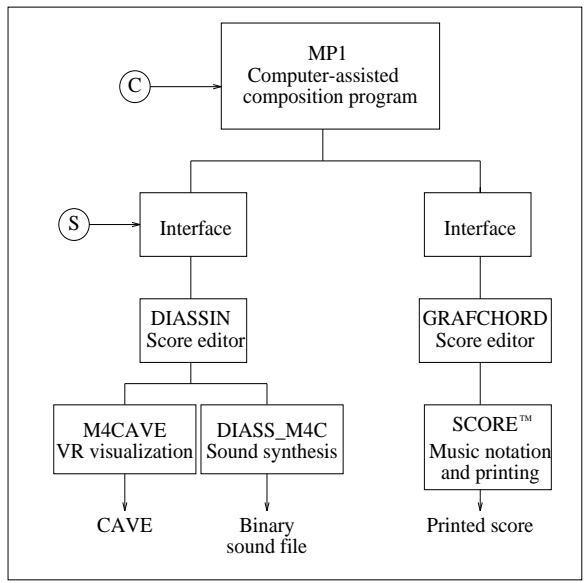,height=3in,width=3in}}
\caption{The
{\em Environment for Music Composition}.
\label{f-env}
}
\end{figure}

Two pieces of software consitute the main tools
of the project: DIASS,
a Digital Instrument for Additive Sound Synthesis,
and M4CAVE,
a program for the visualization of sound objects
in a multimedia environment.
Both are part of a comprehensive
{\em Environment for Music Composition},
which includes additional software for
computer-assisted composition and
automatic music notation.
Figure~\ref{f-env} gives a schematic overview
of the various elements of the {\em Environment\/};
C and S mark the data entry points for
composition and sonification, respectively.

In this section we describe the workings of DIASS;
we will describe M4CAVE after we have discussed
our ideas on scientific sonification.

\subsubsection{The Instrument}
The DIASS instrument functions as part of
the M4C synthesis language developed
by Beauchamp and his associates
at the University of Illinois~\cite{M4C}.
Synthesis languages like M4C are designed
around the notion that the user creates
an instrument together with a score
that references the instrument.
The synthesis program reads the instrument, feeds it
the data from the score file, and computes the final
audio signal, which is then written to a sound file
for later playback~\cite{roads}.

The M4C synthesis language is imbedded in the C language.
As part of the current project, the instrument
and relevant parts of M4C were redesigned
for a distributed-memory environment.
The parallel implementation uses the standard
MPI message-passing library~\cite{MPI}.

Like all additive-synthesis instruments,
DIASS creates sounds through a summation
of simple sine waves.
The basic formula is
\[
  S (t) = \sum_i P_i (t)
  = \sum_i a_i (t) \sin (2 \pi f_i (t) t + \phi_i (t)) .
\]
The individual sine waves that make up a sound
are commonly designated as the ``partials'' of the sound,
hence the symbol $P$.
The sum extends over all partials that are
active at the time $t$;
$a_i$ is the amplitude, $f_i$ the frequency,
and $\phi_i$ the phase of the $i$th partial.
These variables can be modulated periodically or otherwise;
the modulations evolve on a slow time scale,
typically on the order of the duration of a sound.
Phase modulation is barely distinguishable from
frequency modulation, particularly in the case
of time-varying frequency spectra, and is not
implemented in DIASS.

The audible frequencies range roughly
from 20 to 20,000~Hz, although in practice
the upper limit is one-half the sampling frequency
(Nyquist criterion).

The partials in a sound need not be
in any harmonic relationship
(that is, $f_i$ need not be a multiple of some
fundamental frequency $f_0$),
nor do they need to share any other property.
The definition of a sound is purely operational.
What distinguishes one ``sound'' from another
is that certain operations are defined
at the level of a sound and affect all
the partials that make up the sound.

The evolution of a partial can be subject
to many other controls, besides
amplitude and frequency modulation.
Moreover, these controls can affect a single partial
or all the partials in a sound.
For example, reverberation, which represents
the combined effects of the size and
acoustic characteristics of the hall,
affects all the partials in a sound simultaneously,
although not necessarily in the same way.
Furthermore, if a random element is present,
it must be applied at the level of a sound;
otherwise, a complex wave is perceived as
a collection of independent sine waves,
instead of a single sound.
Hence, it is important that all partials
in a sound access the same random number sequence
and that the controls of any partial
that changes its allegiance and moves
from one sound to another be adjusted accordingly.

\begin{table}[htb]
\begin{center}
\caption{Static (S) and dynamic (D) control parameters in DIASS.
 \label{t-controls} }
\vspace*{2ex}
\begin{footnotesize}
\begin{tabular}{|| l | l | l ||}\hline
\multicolumn{1}{||c|}{Level} &
\multicolumn{1}{c|}{Description}&
\multicolumn{1}{c||}{Control Parameter} \\ \hline

Partial & Carrier (sine) wave     & S: Starting time, duration, phase \\
        &                         & D: Amplitude, frequency \\
        & AM (tremolo) wave       & S: Wave type, phase \\
        &                         & D: Amplitude, frequency \\
        & FM (vibrato) wave       & S: Wave type, phase \\
        &                         & D: Amplitude, frequency \\
        & Amplitude transients    & S: Max size \\
        &                         & D: Shape \\
        & Amplitude transient rate& S: Max rate \\
        &                         & D: Rate shape \\
        & Frequency transients    & S: Max size \\
        &                         & D: Shape \\
        & Frequency transient rate& S: Max rate \\
        &                         & D: Rate shape \\
Sound   & Timbre                  & D: Partial-to-sound relation \\
        & Localization            & D: Panning \\
        & Reverberation           & S: Duration, decay rate, mix \\
        & Hall                    & S: Hall size, reflection coefficient \\ \hline
\end{tabular}
\end{footnotesize}
\end{center}
\end{table}
Table~\ref{t-controls} lists the control parameters
that can be applied in DIASS.
Some, like starting time and duration,
do not change for the duration of a sound;
they are static and determined by a single value.
Others are dynamic;
their evolution is controlled by an envelope---a
normalized function consisting of
linear and exponential segments---and a maximum size.
Not all control parameters are totally independent;
some occur only in certain combinations, and
some are designed to reinforce others.

The control parameters give DIASS its flexibility
and make it an instrument suitable for data sonification.
On the other hand, the fact that the control parameters
act at the level of a partial as well as at the level
of a sound (or even at the level of a collection of sounds)
significantly increases its computational complexity.

\subsubsection{The Score}
Input for DIASS consists of a raw score file
detailing the controls.
The raw score file is transformed
into a score file for the instrument---a
collection of ``Instrument cards'' (I-cards),
one for each partial, which are fed
to the instrument by M4C.
The transformation is accomplished
in a number of steps.

Among the controls are certain global operations
(``macros''), which are defined at the level of a sound.
In a first pass, these global controls are expanded
into controls for the individual partials.
The next step consists of the application of
the loudness routines.
These routines operate at the sound level and ensure
that the sounds have the desired loudness.
The final step consists of the application of
the anticlip routines.
For various reasons, historical as well as technical,
sound samples are stored as 16-bit integers.
The anticlip routines guarantee that none of
the sample values produced by the instrument
from the score file exceeds 16 bits.
Because loudness and anticlip play a significant role
in sonification, we discuss the issues in more detail.

\paragraph{Loudness.}
The perception of loudness is a subjective experience.
Although the perceived loudness of a sound is related
to the amplitudes of its constituent partials,
the relation is nonlinear and depends on
the frequencies of the partials.
At the most elementary level,
pure sinusoidal waves of low or high frequencies
require a higher energy flow and therefore a larger
amplitude to achieve the same loudness level
as similar waves at mid-range frequencies.
When waves of different frequencies
are superimposed to form a sound,
the situation becomes still more complicated.
The sum of two tones of the same frequency
produced by two identical instruments
played simultaneously is not perceived as twice
as loud as the tone produced by a single instrument.

An algorithm for data sonification must reflect
these subjective experiences.
For example, when we sonify two degrees of freedom,
mapping one ($x_1$, say) to amplitude and the other
($x_2$, say) to frequency, then we should perceive
equal loudness levels when $x_1$ has the same value, 
irrespective of the values of $x_2$.
Also, when the variable $x_1$ increases or decreases,
we should perceive a proportional increase
or decrease in the loudness level.

The loudness routines in DIASS incorporate
the relevant results of psychoacoustic
research~\cite{montreal}
and give the user full control over the perceived
loudness of a sound.
They also scale each partial so each sample value
fits in a 16-bit register (see Box~2).

\paragraph{Anticlip.}
When several sounds coexist and their waveforms
are added, sample values may exceed 16 bits (overflow),
even when the individual waveforms stay
within the 16-bit limit.
Overflow gives rise to ``clipping''---a popping
noise---when the sound file is played.
The anticlip routines in DIASS check
the score for potential overflow
and rescale the sounds as necessary,
while preserving the ratio of perceived loudness levels.
Thus it is possible to produce an entire sound file
in a single run from the score file, even when
the sounds cover a wide dynamic range.

To appreciate the difficulty inherent in the scaling processes,
consider the case of a sound cluster consisting of
numerous complex sounds, all very loud and resulting in clipping,
followed by a barely audible sound with only two or three partials.
If the cluster's amplitude is brought down to fit the register
capacity, and that of the soft tiny sound following it
is scaled proportionally,
the latter disappears under system noise.
On the other hand, if only the loud cluster is scaled,
the relationship between the two sound events
is completely distorted.
Many times in the past, individual sounds
or groups of sounds were generated separately
and then merged with the help of analog equipment
or an additional digital mixer.
The loudness and anticlip routines in DIASS
deal with this problem by adjusting both loud and
soft sounds so their perceived loudness matches
the desired relationship specified by the user,
and no clipping occurs (see Box~3).

\subsubsection{The Editor}
Features like the loudness routines make DIASS
a fine-tuned, flexible, and precise instrument
suitable for data sonification.
Of course, they require the specification of
significant amounts of input data.
The editor in DIASS is designed to facilitate
this process.
It comes in a ``slow'' and a ``fast'' version.

In the slow version, data are entered
one at a time, either in response to questions
from a menu or through a graphic user interface (GUI).
The process gives the user the opportunity
to build sounds step by step, experiment, and fine-tune
the instrument.
It is suitable for sound composition and for designing
prototype experiments in sonification.
The fast version uses the same code but reads
the responses to the menu questions from a script.
This version is used for sonification experiments.

\subsubsection{Computing Requirements}
The sound synthesis software embodied in DIASS
is computationally intensive (see Box~4).
The instrument proper,
the engine that computes the samples,
has been implemented in a workstation environment
and on the IBM Scalable POWERparallel (SP) system.
Parallelism is implemented at the sound level
to minimize communication among the processors
and enable all partials of a sound to access
the same random number sequence.
In parallel mode, at least four processors are
used---one to distribute the tasks and
supervise the entire run (the ``master'' processor),
a second to mix the results (the ``mixer''),
and at least two ``slave'' nodes to compute
the samples one sound at a time.
Sounds are computed in their starting-time order,
irrespective of their duration or complexity.
(A smart load-balancing algorithm would take into account
the duration of the various sounds and the
number of their partials.)

\begin{center}
\begin{figure}[htbp]
\hspace*{-0.2in}
\centering{\psfig{file=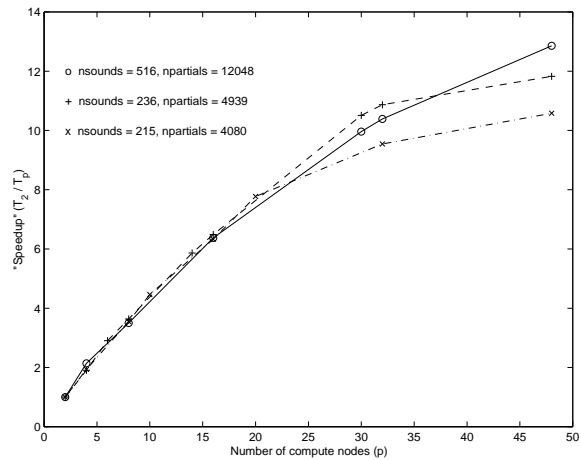,width=3.0in}}
\caption{Timing results for DIASS on an IBM SP.
\label{f-speedup}
}
\end{figure}
\end{center}

Performance depends greatly on the complexity
of the sounds---that is, on the number of partials per sound
and the number of active controls for each partial.
Typically, the time to generate a two-channel sound file
for a 2'26" musical composition
with 236 sounds and 4939 partials
ranges from almost two hours on four processors
to about 10 minutes on 34 processors of the SP.
Figure~\ref{f-speedup} gives some indication of
the speedups one observes in a multiprocessing
environment.
The three graphs correspond to three variants
of the same 2'26" piece with different complexity.
The time $T_p$ refers to a computation
on $p+2$ processors ($p$ ``slaves'');
all times are approximate, as they were
extracted from data given by LoadLeveler,
a not very sophisticated timing instrument
for the SP.
Speedup is measured relative to the performance
on four processors (two compute nodes).
One observes the typical linear speedup
until saturation sets in.
The more complex the piece (the more partials),
the later saturation sets in.
\vspace*{-0.3in}

With a sampling rate of 44,100 samples per second
and two-channel output, a sound file occupies
176~KB per second of sound,
so the sound file for the 2'26"~musical composition
takes close to 25.8~MB of memory.

\section{Data Sonification}
Sonification is the faithful rendition of data in sounds.
When the data come from scientific experiments---actual
physical experiments or computational experiments---
we speak of ``scientific sonification.''
Scientific sonification is therefore the analog
of scientific visualization,
where we deal with aural
instead of visual images.
Because sounds can convey significant amounts
of information, sonification has the potential
to increase the bandwidth of the human/computer
interface.
Yet, its use in scientific computing has received
limited attention.
One reason is, of course, that our sense of vision
seems much more dominant than our sense of hearing.
Another important reason is the lack of
a suitable instrument for scientific sonification.
One of the goals of our project is to demonstrate
that, with an instrument like DIASS, one can probe
multidimensional datasets with surgical precision
and uncover structures that may be hidden to the eye.

\subsection{Past Experiments}
An early experiment with scientific sonification
was done by Yeung~\cite{yeung}.
Seven chemical variables were matched with
seven variables of sound:
two with frequency, one each with loudness,
decay, direction, duration, and rest (silence
between sounds).
His test subjects (professional chemists) were
able to understand the different patterns
of sound representations and correctly classify
the chemicals with a 90\% accuracy rate before and
a 98\% accuracy rate after training.
His experiment showed that motivated expert users
can easily adapt to complex auditory displays.

Recently, a successful application of
scientific sonification was reported in
physics by Pereverzev et al.~\cite{nature}.
The authors were able to detect quantum oscillations
between two weakly coupled reservoirs
of superfluid ${}^3$He using sound,
where oscilloscope traces failed
to reveal structure.

Several other experiments reported in the literature
refer to situations where sounds are used in
combination with visual images for data analysis.
Bly~\cite{bly} ran discriminant analysis experiments
using sound and graphics to represent multivariate,
time-varying, and logarithmic data.
Mezrich et al.~\cite{mezrich} used sound and
dynamic graphics to represent
multivariable time series data.
The ``Exvis'' experiment at the
University of Massachusetts at Lowell~\cite{smith}
expanded this work by assigning sonic attributes
to visual icons.
The importance of sound localization is recognized
by ongoing work at NASA-Ames~\cite{wenzel}.
The evaluation of auditory display techniques
is reported extensively at the annual conferences of ICAD,
the International Conference on Auditory Display;
see~\cite{kramer}.
Sound as a component of the human/computer interface
is discussed in~\cite{buxton}.

Most of the attempts described above used MIDI-controlled
synthesizer sounds, which have drastic limitations
in the number and range of their control parameters.
Bargar et al.~\cite{bargar} at the National Center
for Supercomputing Applications (NCSA)
have developed a complex instrument
with interactive capabilities,
which includes the VSS sound server
for the CAVE virtual-reality environment.

\subsection{What We Have Done So Far}
Much of our work so far has been focused on
the development of DIASS~\cite{ICMC92,ICMC95}.
In addition, we have used DIASS for two preliminary
experiments in scientific sonification, one in chemistry,
the other in materials science.

The first experiment used data from Dr.~Jeff Tilson,
a computational chemist at ANL,
who studied the binding of a carbon atom
to a protonated thiophene molecule.
The data represented the difference in
the energy levels before and after the binding
at $128\times128\times128$ mesh points
of a regular computational grid in space.
Because the data were static,
we arbitrarily identified time with
one of the spatial coordinates
and sonified data in planes parallel to this axis.
The time to traverse a plane over its full length
was usually kept at 30 seconds.
In a typical experiment, we assigned a sound to
every other point in the vertical direction,
distributing the frequencies regularly over
a specified frequency range, and used the data in the
horizontal direction to generate amplitude envelopes
for each of the sounds.
Thus, a sound would become louder or softer
as the data increased or decreased, and
the evolution of the loudness distribution
within the ensemble of 64 sounds was an indicator
of the distribution of the energy difference
before and after the reaction in space.
The sound parameters chosen for the representation
of the data varied from one experiment to another.

The second experiment involved data from
a numerical simulation in materials science.
The scientists were interested in patterns of motion of
magnetic flux vortices through a superconducting medium.
The medium was represented by $384\times256$ mesh points
in a rectangular domain.
As the vortices are driven across the domain,
from left to right, by an external force,
they repel each other but are attracted by
regularly or randomly distributed defects
in the material.
In this experiment,
frequency and frequency modulation (vibrato)
were used to represent movement in the plane,
and changes in loudness were connected to
changes in the speed of a vortex.
A traveling window of constant width
was used to capture the motion of a number
of vortices simultaneously.

These investigations are ongoing,
and the results have not been subjected
to rigorous statistical evaluation.
They have merely served to demonstrate
the capabilities of DIASS and
explore various mappings from
the degrees of freedom in the data to
the parameters controlling the sound synthesis process.
Samples can be heard on the Web~\cite{web-sonification}.

\subsection{What We Have Found So Far}
General conclusions are that 
(i) the sounds produced in each experiment
conveyed information about
the qualitative nature of the data,
and (ii) DIASS is a flexible
and sophisticated tool capable of 
rendering subtle variations in the data.

Changes in some control variables,
such as time, frequency, and amplitude,
are immediately recognizable.
Changes in the combination of partials
in a sound, identifiable through its timbre,
can be recognized with some practice.
Some effects are enhanced by modifiers
like reverberation,
amplitude modulation (tremolo), and
frequency modulation (vibrato).
In some instances, a modifier may lump two,
three, or more degrees of freedom together,
like hall size, duration, and acoustic properties
in the case of reverberation.
Through the proper manipulation of reverberation,
loudness, and spectrum, one can create
the illusion of sounds being produced
at arbitrary locations in a room,
even with only two speakers.

Like the eye,
the ear has a very high power of discrimination.
Even a coarse grid,
such as the temperate tuning used in Western music,
includes about 100 identifiable discrete steps over the
frequency range encompassed by a piano keyboard.
Contemporary music, as well as some non-Western
traditional music, successfully uses smaller increments
of a quarter tone or less for a total of some 200 or more
identifiable steps in the audible range.
Equally discriminating power is available
in the realm of timbre.

Sound is an obvious means to identify regularities
in the time domain, both at the microlevel
and on a larger scale,
and to bring out transitions between random states
and periodic happenings.
Most auditory processes are based on the
recognition of time patterns 
(periodic repetitions giving birth to pitch,
amplitude, or frequency modulation;
spectral consistency creating stable timbres
in a complex sound; etc.),
and the ear is highly attuned to detect
such regularities.

Most conceptual problems in scientific sonification
are related to finding suitable mappings between
the space of data and the space of sounds.
Common sense points toward letting the two domains
share the coordinates of physical space-time if
these are relevant and translating
other degrees of freedom in the data
into separate sound parameters.
On the other hand, it may be advantageous
to experiment with alternative mappings.
Sonification software must be sufficiently flexible
that a user can pair different sets of parameters
in the two domains.

Any mapping between data and sound parameters
must allow for redundancies to enable
the exploration of data at different levels
of complexity.
Similar to visualization software,
sonification software must have utilities
for zooming, modifying the audio palette,
switching between visual and aural representation
of parameters, defining time loops,
slowing down or speeding up, and so forth.

Our experiments also showed that DIASS,
at least in its present form, has its limitations.
One limitation concerns the sheer volume of data
in scientific sonification.
While the composition of a musical piece
(the original intent behind DIASS)
typically entails the handling of
a few thousand sounds,
each with a dozen or so partials,
the number of data points in the
computational chemistry experiment
ran into the millions,
a difference of several orders of magnitude.
By the same token, while a typical amplitude envelope
for a partial or sound in a musical composition
involves ten or even fewer segments,
both experiments required envelopes with
well over 100 such segments.
Another difficulty encountered was the fact
that both experiments required sounds
to be accurately located in space.
While panning is very effective in pinpointing the source
on a horizontal line, suggesting the height
of a sound is a major challenge.
We hope that additions to the software
as well as a contemplated eight-speaker system
will help us get closer to a realistic
three-dimensional representation of sounds.
Finally, to become an effective tool for
sonification, DIASS must operate in real time.
All three concerns are being addressed
in the new C++ version of DIASS currently
under development.

\section{Sound Visualization in a VR Environment}
The notion of sound visualization may at first sight
seem incongruous in the context of data sonification.
However, as has been recognized by several researchers,
the structure of a sound is difficult to detect
without proper training,
and any means of aiding the detection process
will enhance the value of data sonification.
Visualizing sounds is one of these means.
In this project we are focusing on
the visualization of sounds in the CAVE,
a room-size virtual-reality (VR) environment~\cite{cave},
and on the ImmersaDesk, a two-dimensional version.

\subsection{M4CAVE -- A Visualization Tool}
The software collectively known as M4CAVE
takes a score file from the sound synthesis program DIASS
and renders the sounds represented by the score
as visual images in a CAVE or ImmersaDesk.
The images are computed on the fly and are made
to correspond exactly to the sounds one hears through
a one-to-one mapping between control parameters
and visual attributes.
The code, which is written in C++,
uses OpenGL for visualizing objects.

\subsubsection{Graphical Representations}
Currently, M4CAVE can represent sounds
as a collection of spheres (or cubes or polyhedra),
as a cloud of confetti-like particles,
or as  a collection of planes.

\begin{center}
\begin{figure}[htb]
\hspace*{0.0in}
\centering{\psfig{file=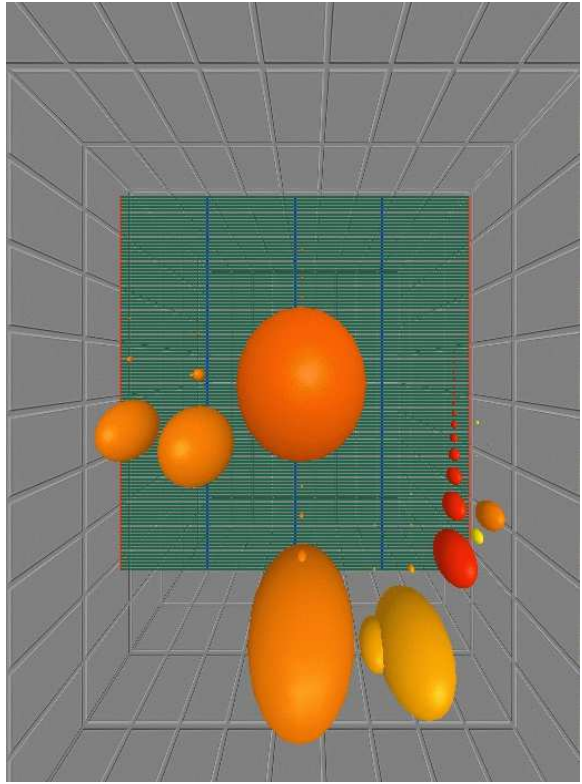,width=3.0in}}
\caption{Visualization of nine sounds.
(Picture taken from a CAVE simulator.)
\label{f-cave}
}
\end{figure}
\end{center}

The spheres representation is the most developed
and incorporates more parameters of a sound
into the visualization than either of the other.
Sounds are visualized as stacks of spheres,
each sphere corresponding to a partial in the sound.
The position of a sphere along the vertical axis
is determined by the frequency of the partial,
and its size is proportional to the amplitude.
A sound's position in the stereo field
determines the placement of the spheres
in the room.
The visual objects rotate or pulse
when tremolo or vibrato is applied,
and their color varies when
reverberation is present.
An optional grid in the background
shows the octaves divided into
twelve equal increments.
Figure~\ref{f-cave}---taken from
our Web site~\cite{web-sonification},
where more samples can be found---shows
a visualization of nine sounds
with different numbers of partials.

\vspace*{-0.5in}
The plane and cloud representations were designed
more on the basis of artistic considerations.
(Remember that the purpose of the visualization
is to aid the perception of sounds.)
The strength of the cloud representation is
in showing tremolo and vibrato in the sound.
The planes representation is unique
in that it limits the visualization to only
one partial (usually the fundamental) of each sound.
The various representations can be combined,
and the mappings chosen for each representation
can be varied by means of a menu.

\subsection{Preliminary Findings}
We have used M4CAVE to explore various mappings
from the sound domain to the visual domain.
Besides the obvious short score files to test
the implementation of these mappings,
we have used score files generated with DIASS
of various musical compositions, notably the
``A.N.L.-folds'' of Tipei~\cite{folds}.
A.N.L.-folds is an example of a
{\em manifold composition},
described in Box~1.
Each member of A.N.L.-folds lasts exactly 2'26''
and comprises between 200 and 500 sounds of
medium to great complexity.
The picture of Fig.~\ref{f-cave} was taken
from a run of one of these A.N.L.-folds.

The combination of visual images and sounds
provides indeed an extremely powerful tool
for uncovering complicated structures.
Sometimes, the sounds reveal features
that are hidden to the eye;
at other times, the visual images illuminate
features that are not easily detectable in the sound.
The two modes of perception reinforce each other,
and both improve with practice.

\section{Larger Issues}
This project is unusual in several respects.
It is somewhat speculative, in the sense that
we don't have much experience with the use of
sound in scientific computing.
This is the main reason why the involvement of
someone expert in the intricacies of the sound world
is critical for its success.
In our case, the expertise comes from the realm
of music composition.

When do we declare ``success''?
Can we reasonably expect that sonification
will evolve to the same level of usefulness
as visualization for computational science?
The answers to these questions depends
on one's expectations.
Ours is a visually oriented culture
{\em par excellence}, and as a society
we watch rather than listen.
Contemporary musical culture is often reduced
to entertainment genres that use a simple-minded
vocabulary---no small impediment to discover
the potential benefits of the world of sound.
But given unusual and unexpected sonorities,
we may yet discover that
we have not lost the ability to listen.

When we engage in this type of research,
it is easy to get swept up by unreasonable
expectations, looking for the ``killer application.''
But the killer application is a phantom,
not worth pursuing.
What we can offer is a systematic investigation
of the potential of a new tool.
If it helps us understand some computational data sets
a little better, or if it enables us
to explore these data sets more easily and in more detail,
we have good reason to claim success.
If the project adds to our understanding
of aural semiotics, we have even more reason
to claim success.
And if none of these successes materializes,
we can still claim that the people involved,
both scientists and musicians,
gained by becoming more familiar with
each other's work and ways of thinking.
Such a rapprochement has, in fact, already
occurred and led to a new ``Discovery'' course
entitled
{\em Music, Science, and Technology}
at UIUC, where some of the issues presented here
are being discussed in a formal educational context.

\section*{Acknowledgments}
This work was partially supported by the
Mathematical, Information, and Computational Sciences Division
subprogram of the Office of Computational and Technology Research,
U.S. Department of Energy, under Contract W-31-109-Eng-38.

\newpage

{\bf Hans G. Kaper}
is Sr.\ Mathematician at Argonne National Laboratory.
After receiving his Ph.D.\ in mathematics from the
University of Groningen (the Netherlands) in 1965,
he held positions at the University of Groningen
and Stanford University.
In 1969 he joined the staff of Argonne.
He was director of the
Mathematics and Computer Science Division
from 1987 to 1991.
Kaper's professional interests are in
applied mathematics, particularly
mathematics of physical systems and
scientific computing.
He is a corresponding member of the
Royal Netherlands Academy of Sciences.
His main interest outside mathematics is classical music.
He is chairman of ``Arts at Argonne,''
concert impresario, and an accomplished pianist.
He can be reached at kaper@mcs.anl.gov or
http://www.mcs.anl.gov/\~{ }kaper/index.html.

{\bf Sever Tipei}
is professor of composition and music theory at the
University of Illinois at Urbana-Champaign (UIUC),
where he also manages the Computer Music Project
of the UIUC Experimental Music Studios.
He has a Diploma in piano 
from the Bucharest Conservatory in Romania
and a DMA in composition from the University of Michigan.
Tipei has been involved in computer music since 1973
and regards the composition of music both as an experimental 
and as a speculative endeavor.
He can be reached at s-tipei@uiuc.edu or
http://cmp-rs.music.uiuc.edu/people/tipei/index.html.

{\bf Elizabeth Wiebel}
was a participant in the
Student Research Participation Program,
which is sponsored by the
Division of Educational Programs
of Argonne National Laboratory.
She is an undergraduate student
at St.\ Norbert College in De Pere, Wisconsin,
where she is pursuing a degree in
mathematics and computer science.
In addition, Wiebel studies and teaches piano
through the St.\ Norbert College music department.
She is currently spending a semester
at Richmond College in London (England).
She can be reached at wiebep@sncac.snc.edu,
F300398@Richmond.ac.uk,
or
http://members.tripod.com/\~{ }LibbyW.

\newpage

\section*{Box~1. \quad Computer-Assisted Music Composition}
The idea of using computers for music composition
goes back to the 1950s, when Lejaren Hiller performed
his experiments at the University of Illinois~\cite{hiller-book}.
The premiere of his Quartet No.\ 4 for strings
``Illiac Suite''~\cite{hiller-quartet}
(May 1957) is generally regarded as
the birth of computer music.
Since then, computers have helped many composers
to algorithmically synthesize new sounds and
produce new pieces for acoustic as well as
digital instruments.
The proceedings of the annual conferences
sponsored by the ICMA
(International Computer Music Association)
are good sources of references~\cite{ICMC}.

Why would a composer need computer assistance 
when composing?
A quick answer is that, as in many other areas,
routine operations can be relegated to the machine.
A more sophisticated reason may be that the composer
may rely on expert systems to write Bach-like chorales
or imitate the mannerisms of Chopin or Rachmaninov.  
There are, however, more compelling reasons
when composing is viewed as a speculative 
and experimental endeavor, rather than as
an ability to manufacture pleasing sounds~\cite{tipei}.

Music is basically a dynamic event evolving
in a multidimensional space;
as such, it can be formalized~\cite{xenakis}.
The composer controls the evolution by supplying
a set of rules, and accepts the output as long as
it is consistent with the logic of the program
and the input data.
If the set of rules allows for a certain degree
of randomness, the output will be different
every time a new ``seed'' is introduced.
The same code and input data may thus produce
an unlimited number of compositions,
all belonging to the same ``equivalence class''
or {\em manifold composition}~\cite{manif}.
The members of a manifold composition are variants
of the same piece; they share the same structure
and are the result of the same process, but differ
in the way specific events are arranged in time.

A nontraditional way of composing,
the manifolds show how high-performace computing
provides the composer with new means
to try out compositional strategies
or materials and hear the results
in a reasonable amount of time.  

\newpage

\section*{Box~2. \quad Loudness}
Sound is transmitted through sound waves---periodic
pressure variations that cause the eardrums to vibrate.
But the perception of loudness has as much to do with
the amount of energy that is carried by the sound wave
as with the processing of this energy that takes place
in the ear and the brain once the sound wave has hit
the eardrums.
The latter is a much more subjective part of the experience.
The algorithms underlying the loudness routines
of DIASS incorporate therefore formal definitions,
as well as results of psychoacoustic research experiments.
We summarize the most relevant elements of the algorithm,
referring the reader to~\cite{roederer} or~\cite{rossing}
for details.

The definition of (perceived) loudness begins
with the consideration of the energy carried
by the sound wave.
The {\em intensity\/} $I$ of a pure tone (sinusoidal sound)
is expressed in terms of its average pressure variation
$\Delta p$ (measured in newton/m$^2$),
\[
  I = 20 \times \log_{10} (\Delta p / \Delta p_0) .
\]
$\Delta p_0$ is a reference value,
usually identified with a traveling wave
of 1,000~Hz at the threshold of hearing,
$\Delta p_0 = 2 \times 10^{-5}$~newton/m$^2$.
The unit of $I$ is the decibel (dB).

Because of the way acoustical vibrations are processed 
in the cochlea (the internal ear),
the sensation of loudness is strongly frequency dependent.
For instance, while an intensity of 50~dB at 1,000~Hz is considered
{\em piano}, the same intensity is barely audible at 60~Hz.
In other words, to produce a given loudness sensation
at low frequencies, a much higher intensity (energy flow)
is needed than at 1,000~Hz.
The intensity $I$ is therefore not a good measure of loudness
if different frequencies are involved.

\begin{figure}[htb]
\hspace*{0.0in}
\centering{\psfig{file=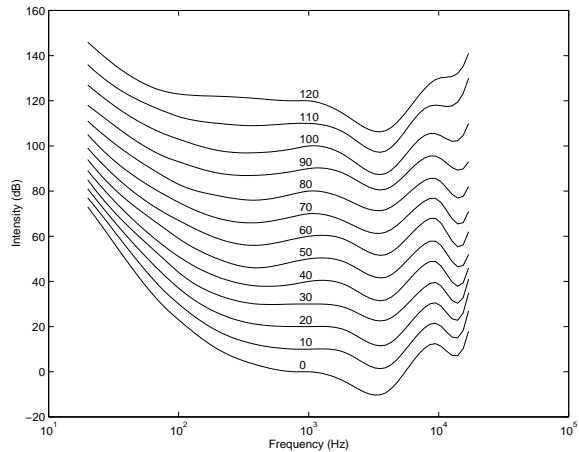,width=3.0in}}
\caption{
Curves of equal loudness (marked in phons)
in the frequency vs.\ intensity plane.
\label{f-loudness}
}
\end{figure}

In the 1930s, Fletcher and Munson~\cite{fletcher-munson}
performed a series of loudness-matching experiments,
from which they derived a set of
{\em curves of equal loudness}.
These are curves in the
frequency ($f$) vs.\ intensity ($I$) plane;
points on the same curve represent
single continuously sounding pure tones
that are perceived as being ``equally loud.''
They are similar to those recommended
by the International Organization for
Standardization (ISO)~\cite{ISO}
and are presented in Fig.~\ref{f-loudness}.
The curves show clearly that,
in order to be perceived as equally loud,
very low and very high frequencies require
much higher intensities (energy)
than frequencies in the middle range
of the spectrum of audible sounds.

The (physical) {\em loudness level\/} $L_p$
of a Fletcher-Munson curve is identified
with the value of $I$
at the reference frequency of 1,000~Hz.
The unit of $L_p$ is the phon.
The Fletcher-Munson curves range from a loudness
level of 0 to 120~phons over a frequency
range from 25 to 16,000~Hz.

The loudness level $L_p$ still does not measure loudness
in an absolute manner: a tone whose $L_p$ is twice
as large does not sound twice as loud.
Following Rossing~\cite{rossing},
we define the (subjective) {\em loudness level\/} $L_s$
in terms of $L_p$ by the formula
$L_s = 2^{(L_p-40)/10}$.
The unit of $L_s$ is a sone.
To be effective, loudness scaling must be done
on the basis of sones.

The loudness of a sound that is composed of
several partials depends on how well the
frequencies of the partials are separated.
With each frequency $f$ is associated
a {\em critical band}, whose width $\Delta f$
is approximely given by the expression~\cite{zwicker-80}
\[
  \Delta f \approx 25 + 75 \left(1 + 1.4(f/1000)^2 \right)^{0.69} .
\]
Intensities within a critical band are added,
and the loudness of a critical band can again
be read off from the Fletcher-Munson tables.
If the frequencies of its constituent partials
are spread over several critical bands,
the loudness of a sound is computed
in accordance with a formula
due to Rossing~\cite{rossing},
\[
  L_s = L_{s,m} + 0.3 \sum_i L_{s,i} .
\]
Here, $L_{s,m}$ is the loudness of the loudest critical band,
and the sum extends over the remaining bands.

The loudness routines in DIASS use
critical band information and
a table derived from the Fletcher-Munson curves
to create complex sounds of specified loudness.

\newpage

\section*{Box~3. \quad Loudness of Sound Clusters}

\begin{figure}[htb]
\hspace*{0.0in}
\centering{\psfig{file=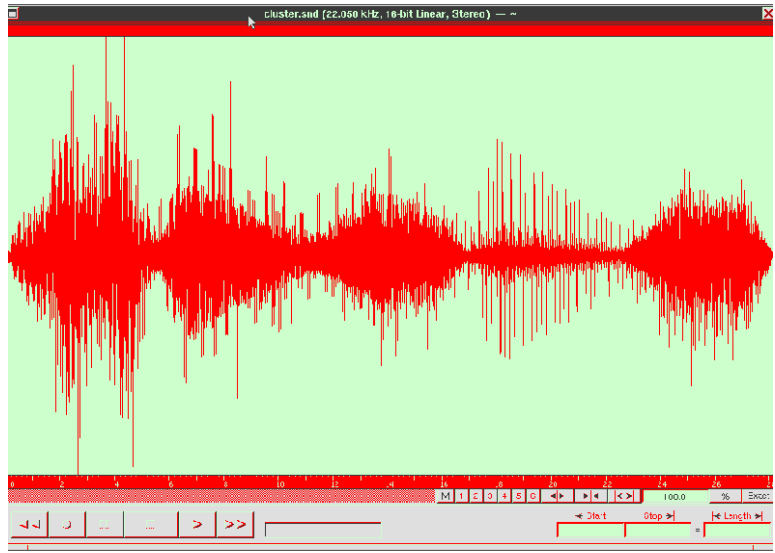,width=4.0in}}
\caption{
Waveform of five sound clusters of equal perceived loudness.
\label{f-clusters}
}
\end{figure}

The waveform of Fig.~\ref{f-clusters},
which was produced with DIASS,
illustrates the concept of equal loudness
across the frequency spectrum and for
different timbres.
The waveform represents five sound clusters,
each lasting 5.5 seconds (except the fourth,
which lasts 5.7 seconds).
The clusters, although of widely different structure,
have been designed to be perceived at the same
loudness level ($2^5$ sones).

The distribution of the sounds within each cluster
is represented schematically in the diagram of
Table~\ref{t-clusters}.
The first sound cluster has 24 sounds.
The fundamental frequencies of the sounds
range from 40 to 5,000 Hz.
Each sound is harmonically tuned;
that is, it is made up of a fundamental
and all its harmonics
(partials whose frequencies
are integer multiples of
the fundamental frequency).
The frequencies are limited to
one-half of the sampling rate
(Nyquist criterion);
hence, the number of partials
in this cluster is 754
(at a sampling rate of 22,050 Hz).
The second sound cluster has 5 sounds,
harmonically tuned, with fundamental frequencies
ranging from 40 to 4,000 Hz;
the number of partials is 113.
The third, fourth, and fifth cluster have
15, 1, and 10 sounds, with 453, 60, and 250 partials,
respectively.
All partials are assigned the same amplitude,
which presents the worst-case scenario
when one tries to obtain the same
perceived loudness for all clusters.

\begin{center}
\begin{table}[h]
\caption{Distribution of fundamentals
in the clusters of Figure~5.  \label{t-clusters}}
\vspace*{1ex}
\begin{footnotesize}
\hspace*{-3em}
\begin{tabular}{||r| c c c c c||}\hline
\multicolumn{1}{||c |}{Fundamental}
         & 24 Sounds   & \hspace{-2em}5 Sounds      & \hspace{-2em}15 Sounds
 & \hspace{-2em}1 Sound       & \hspace{-2em}10 Sounds \\
\multicolumn{1}{||c |}{Frequency}
         &\hspace{-2em}(754 partials) &\hspace{-2em}(113 partials) &\hspace{-2em}(453 partials) &\hspace{-2em}(60 partials)  &\hspace{-2em}(250 partials) \\\hline
         &               &               &               &               & \\
5,000 Hz & \Yes          &               &               &               & \\
4,500 Hz & \Yes          &               &               &               & \Yesm\\
4,000 Hz & \Yes          & \Yesm         & \Yesm         &               & \\
3,000 Hz & \Yes          &               & \Yesm         &               & \\
2,666 Hz & \Yes          &               & \Yesm         &               & \\
2,000 Hz & \Yes          &               &               &               & \Yesm\\
1,666 Hz & \Yes          & \Yesm         & \Yesm         &               & \\
1,333 Hz & \Yes          &               &               &               & \Yesm\\
1,000 Hz & \Yes          &               & \Yesm         &               & \Yesm\\
750 Hz   & \Yes          & \Yesm         & \Yesm         &               & \\
625 Hz   & \Yes          &               &               &               & \Yesm\\
500 Hz   & \Yes          &               & \Yesm         &               & \\
400 Hz   & \Yes          &               &               &               & \Yesm\\
300 Hz   & \Yes          & \Yesm         & \Yesm         &               & \\
200 Hz   & \Yes          &               & \Yesm         &               & \\
165 Hz   & \Yes          &               &               &               & \Yesm\\
130 Hz   & \Yes          &               & \Yesm         &               & \\
90 Hz    & \Yes          &               & \Yesm         &               & \\
80 Hz    & \Yes          &               &               &               & \Yesm\\
70 Hz    & \Yes          &               &               &               & \Yesm\\
60 Hz    & \Yes          &               & \Yesm         &               & \\
53 Hz    & \Yes          &               & \Yesm         &               & \\
46 Hz    & \Yes          &               & \Yesm         &               & \\
40 Hz    & \Yes          & \Yesm         & \Yesm         & \Yesm         & \Yesm\\
         &               &               &               &               & \\\hline
Time     &
\multicolumn{1}{l}{0.0"}&
\multicolumn{1}{l}{\hspace{-1em}5.5"}&
\multicolumn{1}{l}{\hspace{-2em}11.0"}&
\multicolumn{1}{l}{\hspace{-1em}16.5"}&
\multicolumn{1}{l||}{\hspace{-2em}22.2"}\\\hline
\end{tabular}
\end{footnotesize}
\end{table}
\end{center}

\newpage

\section*{Box~4. \quad Computational Complexity}

To give some idea of the computational complexity,
consider the following simple scenario,
where we wish to sonify time-varying data
representing the values of two primary and
several secondary observables measured
over the course of an experiment.
A natural choice is to map the primary observables
onto loudness and frequency and to use
amplitude and frequency modulation to monitor
the secondary observables.
The sample values of the sound wave $S$ must be
calculated from an expression of the form
\be
  S (t) = a (t) \sin \left(2\pi f(t) t + \phi \right) .
  \Label{S}
\ee
The frequency $f$ represents three degrees of freedom:
the carrier frequency $f^C$, and the amplitude
$a^{FM}$ and frequency $f^{FM}$ of the modulating wave,
\be
  f (t)
  =
  f^C (t)
  + a^{FM} (t) \sin \left(2\pi f^{FM} t + \phi^{FM} \right) .
  \Label{f}
\ee
The carrier frequency is identified with a primary observable,
each of the remaining two degrees of freedom can be identified
with a secondary observable,

Similarly, the amplitude $a$ is given by an expression
of the form
\be
  a (t)
  =
  a^C (t)
  + a^{AM} (t) \sin \left(2\pi f^{AM} t + \phi^{AM} \right) .
  \Label{a}
\ee
We compute the carrier amplitude $a^C$
from the observed loudness, which is identified with
one of the (primary) observables, so its value is given.
The amplitude $a^{AM}$ and frequency $f^{AM}$ of the modulation
represent two more degrees of freedom,
which can be identified with
two other secondary observables.
In total, we have therefore two primary and four secondary variables
(not counting the phases, which we assume to be static).

The amplitude $a^C (t)$ must be computed such that
$S (t)$ has the perceived loudness level $L_s (t)$,
\be
  L_s (S(t)) = L_s (t) .
  \Label{L}
\ee
The loudness function $L_s$ is a nonlinear function of
the amplitude and frequency of the partial (sound).
Its computation is done in the loudness routines of DIASS
and involves a significant number of operations,
including table lookups; see Box~2.

On the basis of these formulas we can obtain
a rough estimate of the number of operations
(additions, multiplications,
function evaluations---sine,
exponential, or logarithm,
and table lookups)
required for the computation of a single sample value.
The contribution that is most difficult to estimate is
the computation of the carrier amplitude from
the loudness;
the data in Table~\ref{t-ops}
represent the minimum number of operations.
\begin{table}[htb]
\begin{center}
\caption{Number of operations per partial per sample value.  \label{t-ops} }
\vspace*{2ex}
\begin{small}
\begin{tabular}{|| c || c | c | c | c ||}\hline
Eq.       & Adds & Mults & Fn Evals & Tbl Lkups \\\hline
(\ref{S}) & 1    & 3     & 1        & -           \\
(\ref{f})& 2    & 3     & 1        & -           \\
(\ref{a}) & 2    & 3     & 1        & -           \\
(\ref{L}) & 1    & 3     & 2        & 1           \\\hline
Total     & 6    & 12    & 5        & 1           \\\hline
\end{tabular}
\end{small}
\end{center}
\end{table}
Ignoring phases and so forth, we find a total of
at least 24 operations.
Hence, at the standard rate of 44,100 samples per second,
one needs to perform more than
1.1 million operations per second.

The simultaneous sonification of more observables
is obviously much more complicated;
in fact, the complications grow exponentially.
A careful estimate of the computational complexity
requires an analysis of the anticlip routines,
which is beyond the scope of the present article.


\begin{thebibliography}{99}

\bibitem{buxton}
Baecker, R.~M., J.~Grudin, W.~Buxton, and S.~Greenberg,
\textsl{Readings in Human-Computer Interaction:
Toward the Year 2000},
second edition, Morgan Kaufmann Publ., Inc., San Francisco, 1995

\bibitem{bargar}
Bargar, R., I.~Choi, S.~Das, and C.~Goudeseune,
``Model-based interactive sound for an immersive virtual environment,''
\textsl{Proc.\ 1994 Int'l.\ Computer Music Conference}
(Tokyo, Japan), pp.\ 471--477.

\bibitem{M4C}
Beauchamp, J.,
\textsl{Music 4C Introduction},
Computer Music Project, School of Music,
University of Illinois at Urbana-Champaign, 1993.
URL: http://cmp-rs.music.uiuc.edu/cmp/software/m4c.html

\bibitem{bly}
Bly, S.,
\textsl{Sound and Computer Information Presentation},
Ph.D.\ thesis, University of California -- Davis, 1982
(unpublished)

\bibitem{fletcher-munson}
Fletcher, H.\ and W.~A.~Munson,
``Loudness, its definition, measurement, and calculation,''
{\em J.~Acoust.\ Soc.\ Am.} {\bf 5} (1933), 82

\bibitem{MPI}
Gropp, W., E.~Lusk, and A.~Skjellum,
\textsl{Using MPI: Portable Parallel Programming with the
Message-Passing Interface},
MIT Press, 1994.
See also URL:
http://www.mcs.anl.gov/mpi/index.html

\bibitem{hiller-book}
Hiller, L.\ and L.~Isaacson,
\textsl{Experimental Music},
McGraw-Hill, 1959;
reprinted by Greenwood Press, 1983

\bibitem{hiller-quartet}
Hiller, L.,
\textsl{Computer Music Retrospective},
Compact disc WER 60128-50,
WERGO Schallplatten GmbH,
Mainz, Germany, 1989

\bibitem{ISO}
International Organization for Standardization (ISO),
``Acoustics -- Normal equal-loudness level contours,''
Publ.\ No.~226:1987

\bibitem{ICMC95}
Kaper, H.~G., D.~Ralley, J.~M.~Restrepo, and S.~Tipei,
``Additive synthesis with DIASS\_M4C
on Argonne National Laboratory's IBM POWERparallel System (SP),''
\textsl{Proc.\ 1995 Int'l.\ Computer Music Conference}
(Banff, Canada), pp.\ 351--352

\bibitem{montreal}
Kaper, H.~G., D.~Ralley, and S.~Tipei,
``Perceived equal loudness of complex tones:
A software implementation for computer music composition,''
\textsl{Proc.\ 1996 Int'l.\ Conference in Music Perception and Cognition}
(Montreal, Canada), pp.\ 127--132

\bibitem{kramer}
Kramer, G.\ (ed.),
\textsl{Auditory Display: Sonification, Audification,
and Auditory Interfaces},
Proc.\ ICAD '92,
Addison-Wesley Publ.\ Co., 1994.
For proceedings of later conferences,
consult URL: http://www.santafe.edu/\~{ }icad

\bibitem{ICMC92}
Kriese, C.\ and S.~Tipei,
``A compositional approach to additive synthesis on supercomputers,''
\textsl{Proc.\ 1992 Int'l.\ Computer Music Conference}
(San Jose, California), pp.\ 394--395

\bibitem{mezrich}
Mezrich, J., S.~Frysinger, and R.~Slivjanovski,
``Dynamic representation of multivariate time series data,''
{\em J.~Amer.\ Stat.\ Ass.} {\bf 79} (1984), 34--40

\bibitem{nature}
Pereverzev, S.~V., A.~Loshak, S.~Backhaus, J.~C.~Davis,
and R.~E.~Packard,
``Quantum oscillations between two weakly coupled
reservoirs of superfluid ${}^3$He,''
{\em Nature} {\bf 388} (1997), 449-451

\bibitem{roads}
Roads, C.,
\textsl{The Computer Music Tutorial},
MIT Press, Cambridge, Mass., 1996

\bibitem{roederer}
Roederer, J.~G.,
\textsl{The Physics and Psychophysics of Music},
3rd edition.
Springer-Verlag, 1995

\bibitem{rossing}
Rossing, T.~D.,
\textsl{The Science of Sound},
Addison-Wesley Publ.\ Co., 1990

\bibitem{ICMC}
Simoni, M.\ (ed.),
\textsl{Proc.\ ICMC98, Int'l Computer Music Conference}
(Ann Arbor, Michigan), October 1998;
see also proceedings of earlier conferences

\bibitem{smith}
Smith, S.\ and M.~Williams,
``The use of sound in an exploratory visualization experiment,''
CS Dept., U.~Mass. at Lowell,
tech report R-89-002, 1989

\bibitem{tipei}
Tipei, S.,
``The computer: A composer's collaborator,''
{\em Leonardo\/} {\bf 22}(2) 1989, 189--195

\bibitem{manif}
Tipei, S.,
``Manifold compositions --- A (super)computer-assisted composition
experiment in progress,''
\textsl{Proc.\ 1989 Int'l.\ Computer Music Conference}
(Columbus, Ohio), pp.\ 324--327

\bibitem{folds}
Tipei, S.,
``A.N.L.-folds.\
mani 1943-0000;
mani 1985r-2101;
mani 1943r-0101;
mani 1996m-1001;
mani 1996t-2001''
(1996).
Report ANL/MCS-P679-0897,
Mathematics and Computer Science Division,
Argonne National Laboratory

\bibitem{web-sonification}
URL:
http://mcs.anl.gov/appliedmath/Sonification/index.html

\bibitem{cave}
URL:
http://www.evl.uic.edu/pape/CAVE/prog/CAVEGuide.html

\bibitem{wenzel}
Wenzel, E., S.~Fisher, P.~Stone, and S.~Foster,
``A system for three-dimensional acoustic `visualization'
in a virtual environment workstation,''
\textsl{Proc.\ Visualization '90:
First IEEE Conf.\ on Visualization},
IEEE Computer Society Press, Washington,
pp. 329--337

\bibitem{xenakis}
Xenakis, I.,
\textsl{Formalized Music: Thought and Mathematics
in Musical Composition},
revised edition, Pendragon Press, 1992

\bibitem{yeung}
Yeung, E.,
``Pattern recognition by audio representation
of multivariate analytical data,''
\textit{Analytical Chemistry} {\bf 52} (1980), 1120--1123

\bibitem{zwicker-80}
Zwicker, E.\ and E.~Terhardt,
``Analytical expressions for critical-band rate
and critical bandwidth as a function of frequency,''
{\em J.~Acoust.\ Soc.\ Am.} {\bf 68} (1980), 5

\end{thebibliography}
\end{document}